 \documentclass[smallabstract,smallcaptions]{dccpaper}

\usepackage[utf8]{inputenc} 
\usepackage[T1]{fontenc}    
\usepackage{cite}       

\usepackage{epsfig}
\usepackage{amsmath}
\usepackage{amssymb}
\usepackage{color}
\usepackage{url}

\newlength{\figurewidth}
\newlength{\smallfigurewidth}

\usepackage{float}
\usepackage{graphicx}
\usepackage{subfig}
\usepackage{array}
\usepackage{multirow}
\usepackage{flushend}

\begin{document}

\title
{\large
\textbf{Multiscale Point Cloud Geometry Compression}
}

\author{%
Jianqiang Wang$^\star$, Dandan Ding$^\dag$, Zhu Li$^\ddag$, and Zhan Ma$^\star$\\[0.15em]
$^\star$Nanjing University, $^\dag$Hangzhou Normal University, $^\ddag$University of Missouri\\[0.15em]
}

\maketitle
\thispagestyle{empty}
\begin{abstract}

Recent years have witnessed the growth of point cloud based applications because of its realistic and fine-grained representation of 3D objects and scenes. However, it is a challenging problem to compress sparse, unstructured, and high-precision 3D points for efficient communication. In this paper, leveraging the sparsity nature of point cloud, we propose a multiscale end-to-end learning framework which hierarchically reconstructs the 3D Point Cloud Geometry (PCG) via progressive re-sampling. The framework is developed on top of a sparse convolution based autoencoder for point cloud compression and reconstruction. For the input PCG which has only the binary occupancy attribute, our framework translates it to a downscaled point cloud at the bottleneck layer which possesses both geometry and associated feature attributes. Then, the geometric occupancy is losslessly compressed {using an octree codec} and the feature attributes are lossy compressed {using a learned probabilistic context model.}Compared to state-of-the-art Video-based Point Cloud Compression (V-PCC) and Geometry-based PCC (G-PCC) schemes standardized by the Moving Picture Experts Group (MPEG), our method achieves more than 40\% and 70\% BD-Rate (Bj\o ntegaard Delta Rate) reduction, respectively. Its encoding runtime is comparable to that of G-PCC, which is only 1.5\% of V-PCC.
\end{abstract}

\section{Introduction}
Recently, Point Cloud (PC) has emerged as a promising format to realistically represent the 3D object and scene, by a set of unstructured points sparsely distributed in a 3D space. Similar to 2D images or videos, high-efficiency Point Cloud Compression (PCC) is of great interest for networked application enabling, such as augmented reality, autonomous driving, etc.

Existing representative PCC methodologies were developed under the efforts of the experts in MPEG~\cite{Schwarz2019EmergingMS}. Two popular architectures were adopted, having respective Video-based PCC (V-PCC) and Geometry-based PCC (G-PCC) approaches. The V-PCC utilized the 3D-to-2D projection, by which 2D video codec was applied to code projected planes; While G-PCC relied on 3D models (e.g, octree or triangle surface) for encoding the 3D content directly. More details could be found in MPEG standard specification and its companion paper~\cite{Schwarz2019EmergingMS}. 
These standard compliant PCC methods separately compress geometry and color (or reflectance) attributes. In this work, we will first emphasize the Point Cloud Geometry (PCG) compression.

In contrast to the aforementioned solutions, recent works have attempted to apply the deep learning techniques, e.g., 3D Convolutional Neural Network (CNN) or PointNet based Auto-Encoders (AEs), to compress PCG~\cite{quach2019learning, guarda2019point,wang2019learned,Yan2019DeepAL,huang20193d,Wen2020ICME}. 
In these explorations, Learned-PCGC in~\cite{wang2019learned} has demonstrated very encouraging rate-distortion efficiency, offering a noticeable performance gain over MPEG G-PCC and comparable performance to V-PCC.
{In Learned-PCGC, the PCG is represented using the non-overlapped 3D binary occupancy block for compression. Though most voxels are null, 3D convolutions are still uniformly applied on each voxel without efficiently exploiting the point cloud sparsity. It also demands a large amount of space and computational resources for processing, making it difficult for practical application.}

This paper aims to develop a {more affordable} learning-based approach for high-efficiency PCG compression. 
{Similar to the traditional octree model based PCG codec~\cite{Schnabel2006Octree}, we progressively downscale the input PCG to implicitly embed the original binary occupancy information in a local block using convolutional transforms. 
{And for efficient and low-complexity tensor processing, we use the sparse convolution~\cite{choy20194d} rather than the conventional dense convolution.}
As will be shown in subsequent sections, the progressive scaling could well capture the characteristics of sparse and unstructured points for compact representation.} To this end, we perform a multiscale re-sampling (e.g., downscaling in the encoder, and upscaling in the decoder) to exploit the sparsity and implicitly embed local 3D geometric structural variations into feature attributes of the latent representation (a.k.a., a downsized point cloud from original input PCG with binary occupancy). 
At the bottleneck, we separately process the {geometry and associated features of the latent representation}, e.g., the geometric occupancy is losslessly compressed {using an octree codec} and the feature attributes are lossy compressed {using a learned probabilistic context model.}
Finally, the proposed end-to-end learning scheme is fulfilled by examining the encoded bit rate and the distortions of each {scale} measured by Binary Cross Entropy (BCE) loss.

Using the common test point clouds suggested by MPEG and JPEG, our method outperforms state-of-the-art MPEG V-PCC and G-PCC by more than 40\% and 70\% BD-Rate gains, respectively. Besides, better subjective reconstruction quality is also visualized. 
{When compared with recent Learned-PCGC~\cite{wang2019learned}, our method achieves more than 30\% BD-Rate saving and the complexity consumption is also greatly reduced.}
Since our method leverages the sparse convolution, it can be easily devised to process large-scale point clouds, with well-balanced efficiency and complexity trade-off. Moreover, our multiscale optimization makes the model easier for training with a faster convergence rate.  In contrast to the single-scale approach, 
{our multiscale reconstruction reduces about 75\% memory space requirement, and has $\approx$3$\times$ computational speedup.}
At the end of our experiments, we further conduct serial studies to examine the potential of our method in practice. For example, our approach could receive another 7.2\% BD-Rate reduction when modelling latent elements conditioned on autoregressive priors. 
\section{Related Works} \label{sec:related_work}
In the following, we briefly review the related PCG compression and sparse convolution work. 

{\bf PCG Compression.} 
Octree decomposition~\cite{Schnabel2006Octree} is the most straightforward way to model PCG representation {of various point clouds}.
It is adopted in MPEG G-PCC~\cite{Schwarz2019EmergingMS} known as \textit{octree geometry codec}. 
Alternatively, the surface model is often used for representing dense point cloud~\cite{Schwarz2019EmergingMS}. 
In G-PCC,  triangle meshes or triangle soup (\textit{trisoup}) is used to approximate the surface of 3D model, which is referred to as \textit{trisoup geometry codec}. 
Different from G-PCC which leverages 3D models for point cloud compression, MPEG V-PCC~\cite{Schwarz2019EmergingMS} first performs 3D-to-2D projection, by which successful video codecs, e.g., High-Efficiency Video Coding (HEVC)~\cite{Sullivan2013Overview}, then can be applied to encode the projected planes, depth maps, etc., over the time. Currently, V-PCC offers state-of-the-art efficiency when compressing the dense point clouds.

Recently, with the advancements of deep learning techniques, learning-based PCG compression approaches have emerged. Most of them inherit the network architectures from learned 2D image compression~\cite{minnen2018joint}, but use 3D convolutions instead. For example, Quach \textit{et al.}~\cite{quach2019learning}, Wang \textit{et al.}~\cite{wang2019learned}, and Guarda \textit{et al.}~\cite{guarda2019point} attempted to represent the 3D occupancy model of voxelized PCG using 3D CNN based AEs, 
where the occupancy reconstruction is modelled as a classification problem  via the BCE loss optimization. Among them, Wang \textit{et al.}~\cite{wang2019learned} had reported the leading PCG compression efficiency at the cost of excessive computational complexity in both time and space dimension.
This is because it relies on the volumetric models (e.g., block-by-block processing), without considering the sparsity nature of point cloud.
On the contrary, Yan~\textit{et al.}~\cite{Yan2019DeepAL}, Huang~\textit{et al.}~\cite{huang20193d}, and Wen~\textit{ et al.}~\cite{Wen2020ICME} applied  point-based approaches for compression instead of aforementioned volumetric model. For example,
Huang~\textit{et al.}~\cite{huang20193d} devised a hierarchical AE with multiscale loss {for fine-grained reconstruction.} Wen~\textit{et al.}~\cite{Wen2020ICME} proposed an adaptive octree-guided network for large-scale point cloud support.
In these point-based methods, although redundant calculation can be reduced, only marginal improvement over MPEG G-PCC was achieved because the correlation across voxels are not fully exploited.
To overcome above challenges, in this work, we propose to incorporate the 3D sparse tensors to a point-based processing framework, for both low-complexity and high coding efficiency.

{\bf Sparse Convolution.}
A number of explorations have been made to exploit the point cloud sparsity, such as the octree-based CNN~\cite{wang2017cnn} and sparse CNN~\cite{choy20194d,graham20183d,Graham2018UnsupervisedLW}. For sparse CNN,  data tensor is represented using a set of coordinates $C=\{(x_i, y_i, z_i)\}_i$ and associated features $F=\{f_i\}_i$,
{and the convolution only aggregates features at positively-occupied coordinates. It is defined in~\cite{choy20194d} as:}
\begin{equation}
\vspace{-0.025in}
f_{u}^{out} = \sum\nolimits_{i \in N^{3}(u, C^{in})} W_{i} f_{u+i}^{in} 
\quad\text{for}\quad
u \in C^{out}
\end{equation} 
where $C^{in}$ and $C^{out}$ are input  and output coordinates.
$f_{u}^{in}$ and $f_{u}^{out}$ are input and output feature vectors at coordinate $u$. 
$N^{3}(u, C^{in}) = \{i|u+i \in C^{in}, i\in N^{3}\}$ defines a 3D convolutional kernel, covering a set of locations centered at $u$ with offset $i$'s in $C^{in}$. $W_i$ denotes the kernel value at offset $i$.
This sparse convolution leverages the point cloud sparsity for complexity reduction by applying the computations only on positively-occupied voxels.
Recent explorations~\cite{choy20194d} have shown that sparse CNN is capable of processing large-scale point clouds, with a very promising performance in semantic segmentation, object detection, etc., but it still lacks of studies in point cloud compression.

\section{Multiscale PCG Compression}
\begin{figure*}[t]
	\centering
	\vspace{-0.10in}
	\includegraphics[width=5.5in]{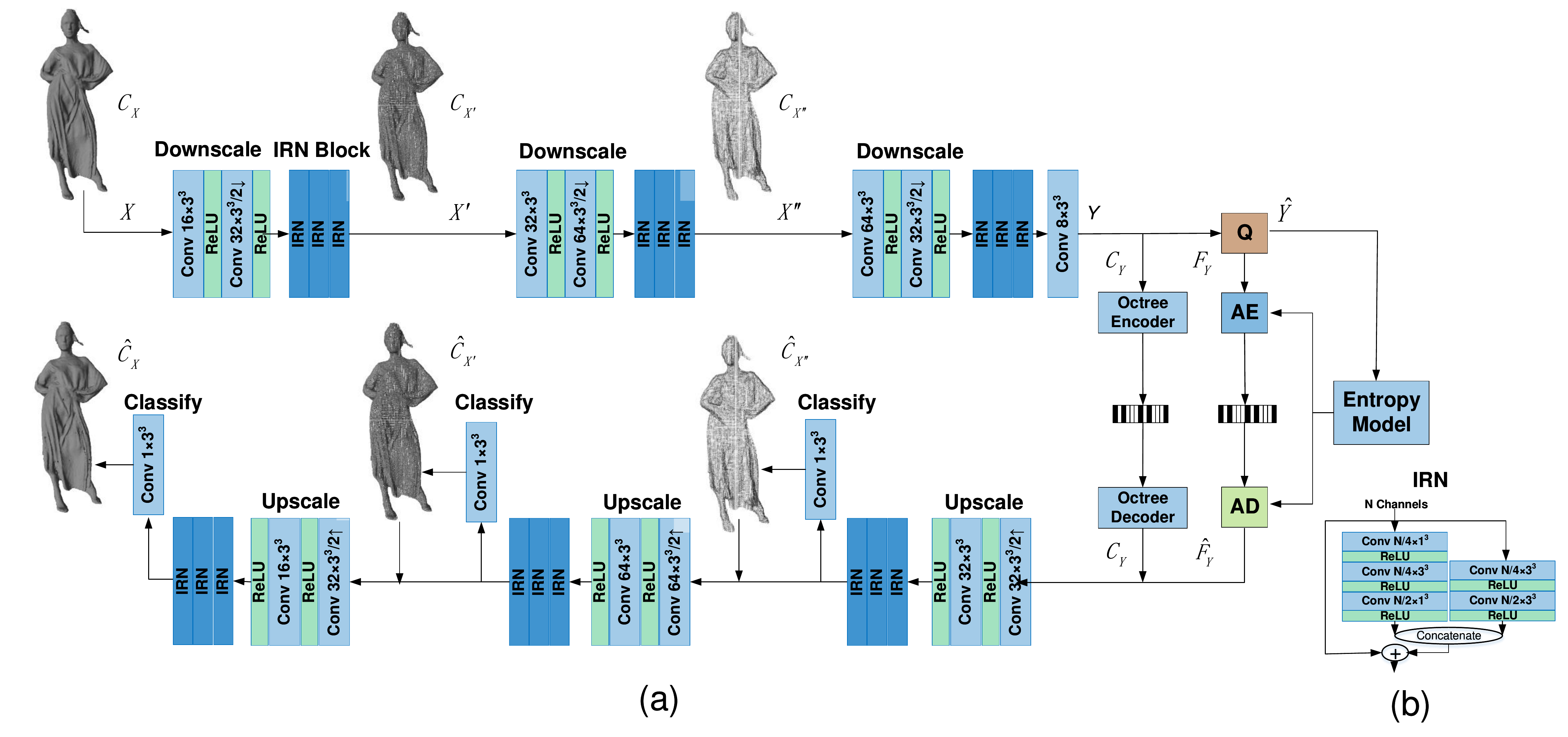}
	\vspace{-0.15in}
	\caption{{\bf Multiscale PCG Compression}.  (a) ``Conv $c\times n^3$'' denotes the sparse convolution with $c$ output channels and $n\times n\times n$ kernel size, ``$s\uparrow$'' and ``$s\downarrow$'' represent upscaling and downscaling at a factor of $s$, and ``ReLU'' stands for the Rectified Linear Unit.  ``Q'' implies ``Quantization'', ``AE'' and ``AD'' are Arithmetic Encoder and Decoder, respectively. ``IRN'' is Inception-Residual Network used for efficient feature aggregation. (b) The network structure of IRN unit.}
	\label{fig:framework}
\vspace{-0.15in}
\end{figure*}

The proposed multiscale PCG compression is illustrated in Fig.~\ref{fig:framework}. It is built upon the popular convolutional AEs to deal with the sparse tensors which are represented using geometry coordinates $C$ (positively occupied) and feature attributes $F$. As such, the input PCG $X$ is a sparse tensor $\{C_X, F_{X}\}$ with all-ones vector $F_{X}$ to indicate the occupancy. In the encoder, $X$ is progressively downsampled to multiple scales (i.e., $X'$, $X''$, and $Y$) by implicitly learning and embedding local structures (a.k.a., the distribution of positively occupied voxels in the volumetric representation of a point cloud) into $F$ components. At the bottleneck, both geometry coordinates $C_{Y}$ and feature attributes $F_Y$ of $Y$ of the latent representation will be encoded into a binary string, {where $C_{Y}$ can be regarded as the structural keypoints describing the coarse skeleton of the input PCG.} Correspondingly, in the decoder, $C_{Y}$ and $\hat{F}_Y$  will be decoded and refined to produce multiscale reconstructions hierarchically.

{\bf Sparse Convolution based Multiscale Resampling.}
We adopt sparse convolutions~\cite{choy20194d} for low-complexity tensor processing. In the meantime, we follow~\cite{wang2019learned} to employ the Inception-Residual Network (IRN)~\cite{Szegedy2017Inceptionv4IA} unit for efficient feature extraction. 
Three IRN blocks, each of which has three consecutive IRN units, are devised after each down-scaling and up-scaling steps.
As for encoder-side downscaling, we apply the convolution with a stride of two to halve the scale of each geometric dimension every step (e.g., $\times\frac{1}{2}$). 
Extensive simulations suggest that repeating the downscaling three times is well justified for compression efficiency. 

\begin{figure*}[t]
	\centering
	\vspace{-0.2in}
	\includegraphics[width=5.5in]{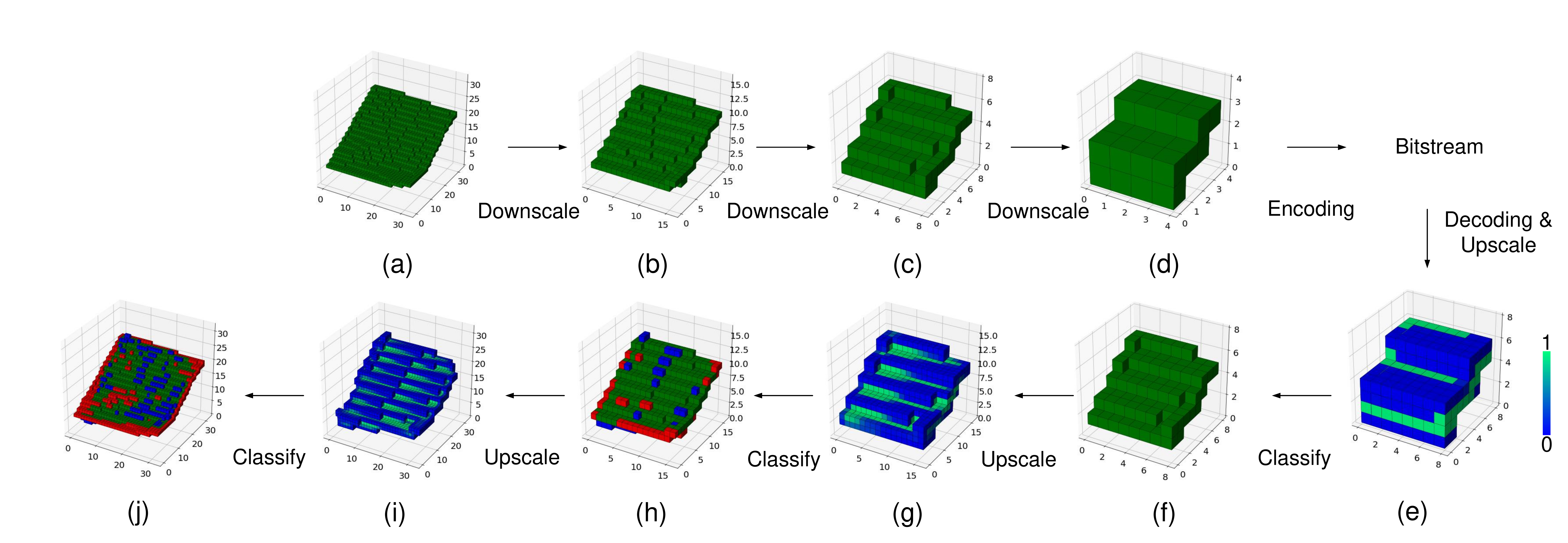}
	\vspace{-0.10in}
	\caption{{\bf Binary classification based hierarchical reconstruction.} 
		The top part shows the encoding process: \textbf{(a)}-\textbf{(d)} are exemplified from a size of $32^3$ to  $4^3$, by halving each geometric dimension scale step-by-step;
		The bottom part illustrates corresponding  hierarchical reconstructions, where geometric models are upscaled and classified gradually from the rightmost to the leftmost position: \textbf{(e)}, \textbf{(g)}, \textbf{(i)} are convolutionally upscaled from lower scales having sizes of  $8^3$, $16^3$ and $32^3$.  Different colors are used to measure the probabilities of voxel-being-occupied (i.e., the greener, the closer  to 1, and the bluer, the closer to 0);	
		\textbf{(f)}, \textbf{(h)}, \textbf{(j)} are the reconstructions after classification, with green blocks for true classified voxels,  blue for false positive, and red for false negative voxels. 
		{In training, all voxels will be used to alleviate errors in the next step, while in inference, only positive ones will be utilized.}}
	\label{fig:reconstruct}
\vspace{-0.15in}
\end{figure*}

The decoding process mirrors the encoder to generate reconstructions at different scales progressively. We use transposed convolution with a stride of two for upscaling (e.g., $\times 2$ in each dimension).  Then we reconstruct the geometry details by pruning false voxels and extracting true occupied voxels using binary classification, which follows a hierarchical, coarse-to-fine refinement as shown in Fig.~\ref{fig:framework}(a) 
and Fig.~\ref{fig:reconstruct}.  More specifically, reconstruction from the preceding lower scale will be upscaled by augmenting decoded feature attributes for finer geometry refinement at the current layer. 

{\bf Binary Classification based Hierarchical Reconstruction.}
Binary classification is applied to classify whether generated voxels are occupied or not. We utilize a sparse convolutional layer to produce the probability of voxel-being-occupied after consecutive convolutions (see Fig.~\ref{fig:framework}(a)). We set this BCE loss in training, i.e.,
\vspace{-0.025in}
\begin{equation}
\label{bceloss}
L_{\rm BCE} = \frac{1}{N}\sum\nolimits_{i}-(x_{i}\log(p_{i}) + (1-x_{i})\log(1 - p_{i})), 
\vspace{-0.025in}
\end{equation} 
where $x_{i}$ is the voxel label that is either truly occupied (1) or empty (0), and $p_{i}$ is the probability of voxel-being-occupied, which is activated by the $\sf sigmoid$ function.
{Alternatively, Mean Square Error (MSE) can also be used to measure distortion when transforming PCG to truncated signed distance fields (TSDF) as in~\cite{Tang2020DeepIV}.}

In the inference stage, the voxel-being-occupied is determined by the value of $p$.  Specifically, we sort $p$ and extract top-$k$ voxels, which are regarded as the voxels most likely to be occupied. 
$k$ can be used to adjust the density of reconstructed PCG. Experimental studies have found that setting $k$ to the number of points in the ground truth label at each scale can achieve almost the lowest point-to-point  distortion.
{A similar observation is also reported in~\cite{wang2019learned}.}

For the hierarchical reconstruction at different scales, the same mechanism is adopted.
And the multiscale loss is comprised of scale-dependent BCE loss, i.e., $D = \frac{1}{M} \sum\nolimits_{j=1}^{M} L_{\rm BCE}^{j}$,
with $j$ as the scale index, and $M$ as the total number of scales.
Figure~\ref{fig:reconstruct} visualizes the hierarchical reconstruction process with step-wise illustrations, where the reconstruction of subsequent scale depends on the results from the proceeding layer.
In this way, the geometry details are generated gradually while the sparsity is leveraged in each stage for reducing the complexity consumption.

{\bf Latent Representation Compression.}
\label{sec:latent}
The latent representation $Y$ is comprised of the positively-occupied coordinates $C_{Y}$ (a.k.a., geometry) and associated feature attributes $F_Y$. In this work, we suggest compressing $C_{Y}$ and $F_Y$ separately.

$C_{Y}$ is losslessly encoded using the octree geometry codec in G-PCC~\cite{tmc13} since it only consumes a very small amount of bits (i.e., around 0.025 bpp). Although applying the lossy compression may further reduce the bits consumption, it does not bring noticeable gains. 
{And more importantly, $C_{Y}$ can be referred to as the skeleton keypoints of input PCG, thus its distortion may lead to severe quality degradation.}

Note that $F_Y$ is derived in progressive re-sampling for characterizing the 3D block occupancy distribution by implicitly embedding local structural correlations into feature components. We propose to perform a uniform quantization on $F_Y$ for rate adaptation prior to entropy coding. 
In inference, we directly quantize $F_{Y}$ using $\sf rounding$, i.e., $\hat{F_{Y}}= \lfloor F_{Y} \rceil $. While in training, we approximate the rounding operation by adding uniform noise to ensure the differentiability in back propogation~\cite{balle2018variational}.
{Afterwards, the quantized $\hat{F_{Y}}$ is encoded by arithmetic coding using a probabilistic model. We first devise  a non-parametric, fully factorized density model~\cite{balle2018variational}, and the estimated probability $p_{\hat{F_{Y}}|\psi}$ is based on ~\textit{factorized prior}, having $\psi$ as its parameters.}
We further estimate $p_{\hat{F_{Y}}}$ using conditional entropy models with the help from the autoregressive neighbors and hyper priors. {We will give more details in subsequent discussions.}

\section{Experimental Results}
\label{sec:exp}
{\bf Training.} 
We use ShapeNet~\cite{chang2015shapenet} for training, which contains $\approx$51,300 CAD surface models. These models are first densely sampled to generate point clouds, and then randomly rotated and quantized to 7-bit precision for each dimension. The number of points in each point cloud is randomized without imposing any constraints. 

In training, we optimize the Lagrangian loss, i.e., $J_{loss} = R + \lambda D$,
where $R$ is the compressed bit rate by encoding the $\hat{F_{Y}}$.
Here, $R_{\hat{F_{Y}}}$ is calculated by the probability density function $p_{\hat{F_{Y}}}$, 
i.e., $R_{\hat{F_{Y}}} =\frac{1}{N} \sum\nolimits_{i}^{N}-\log_{2}(p_{\hat{F_{Y}}_{i}}),$
,$D$ is obtained from the aforementioned multiscale BCE loss. Parameter $\lambda$ controls the rate-distortion trade-off for each individual bit rate. In our experiments, to cover a wide range of bit rate,  we empirically adapt $\lambda$ between 0.25 and 10.
The Adam optimizer is utilized with a learning rate decayed from 0.0008 to 0.00002. We train the model for around 32,000 batches with a batch size of 8. 

{\bf Performance Evaluation.} 
In the tests, we select ten dense point clouds for evaluation, including four models from 8i Voxelized Full Bodies (8iVFB)~\cite{8i20178i} (Longdress, Loot, Redandblack, Soldier), two models from Owlii dynamic human mesh~\cite{xu2017owlii} (Basketball\_player, Dancer), and four models from Microsoft Voxelized Upper Bodies (MVUB)~\cite{microsoft2019microsoft} (Andrew, David, Phil, Sarah). 
These point clouds cover different scales and structures, which are adopted either in MPEG Common Test Condition (CTC)~\cite{Sebastian2018common} or in JPEG Pleno  CTC~\cite{JPEG2019CTC} for compression task exploration.
{Extending our approach to sparse point clouds, such as the LiDAR point clouds is an interesting topic for our future study.}
%

\begin{figure*}[thbp]
\vspace{-0.1in}
\centering
\subfloat{\includegraphics[width=1.42in]{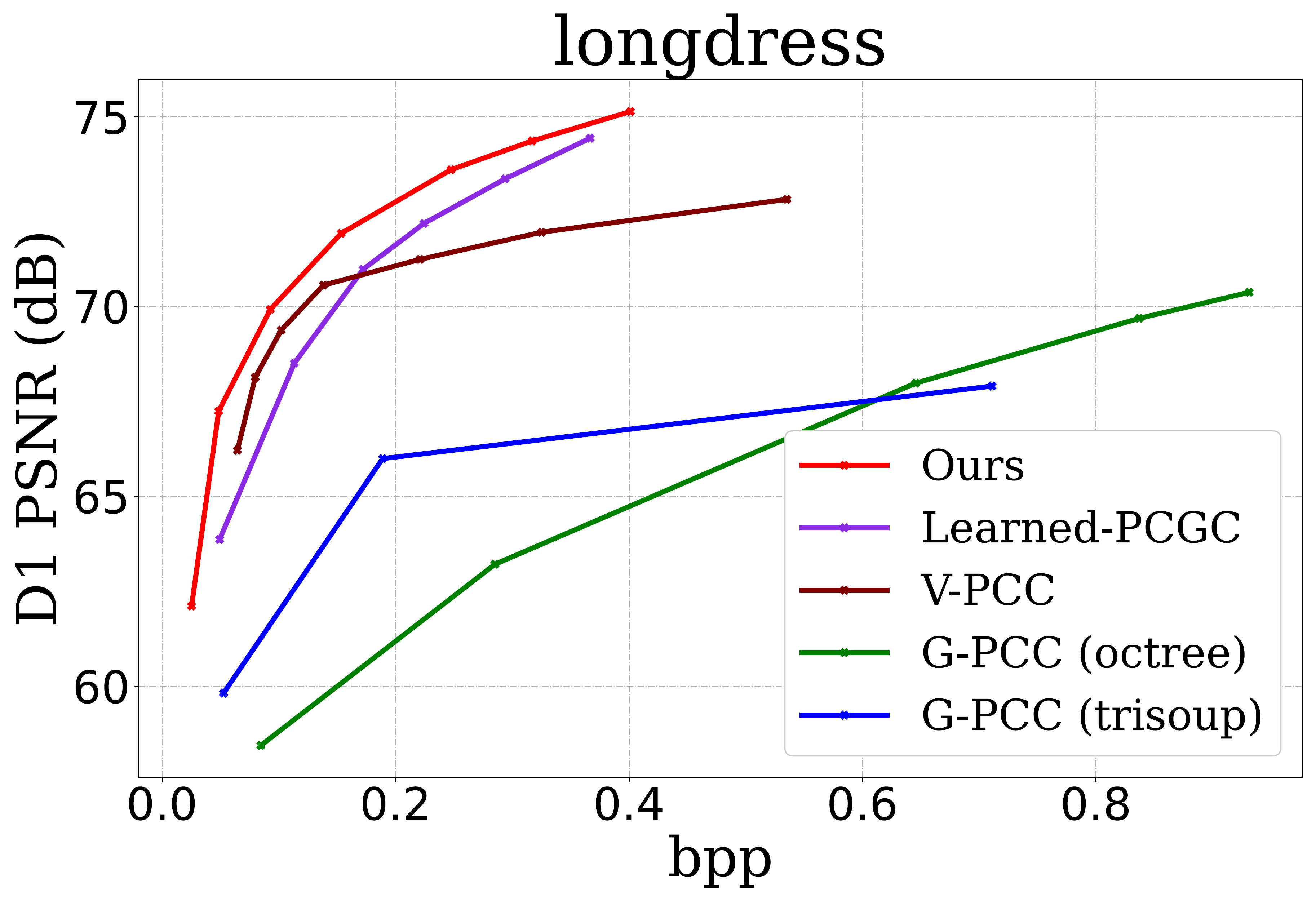}}%
\subfloat{\includegraphics[width=1.42in]{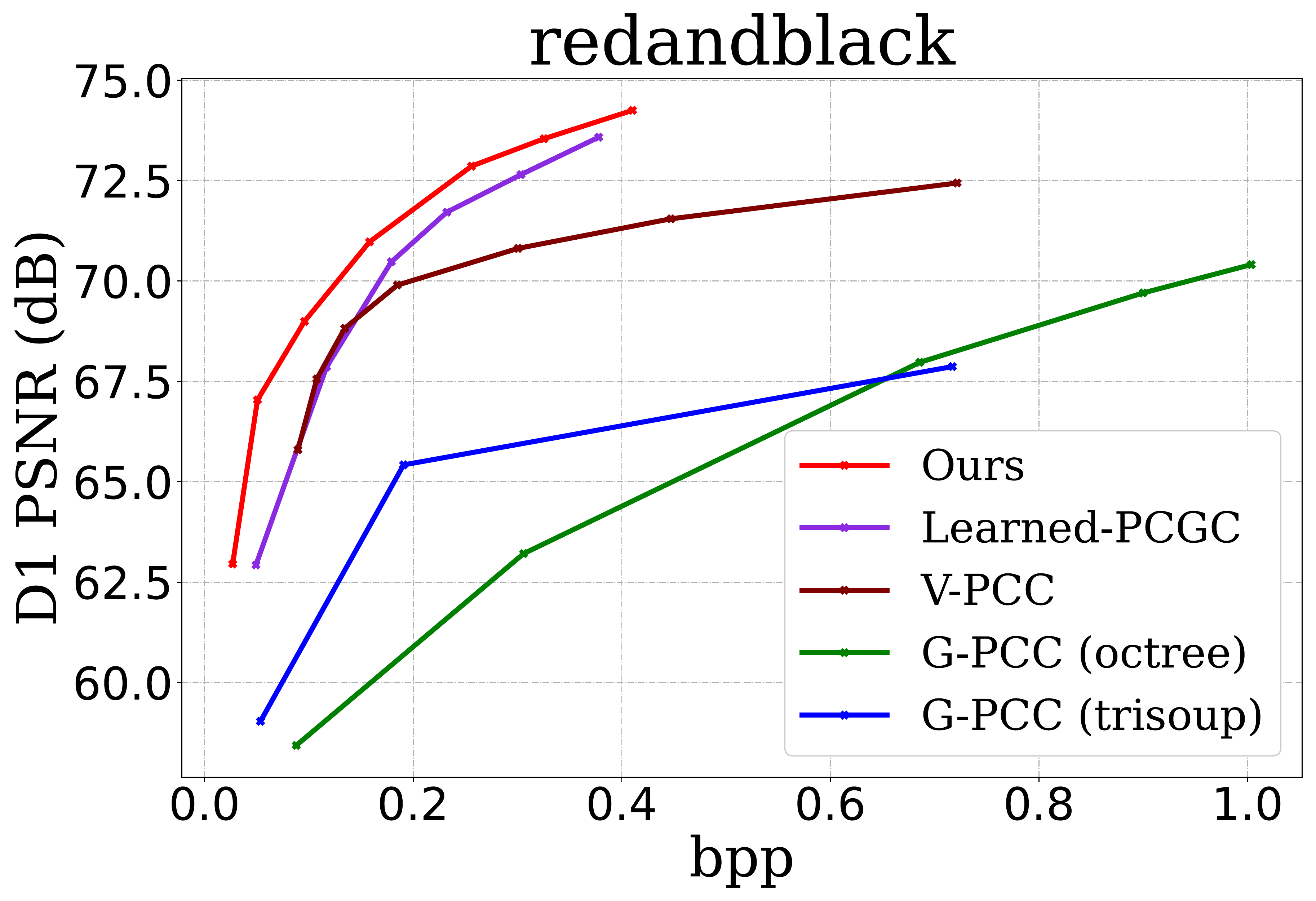}}
\subfloat{\includegraphics[width=1.42in]{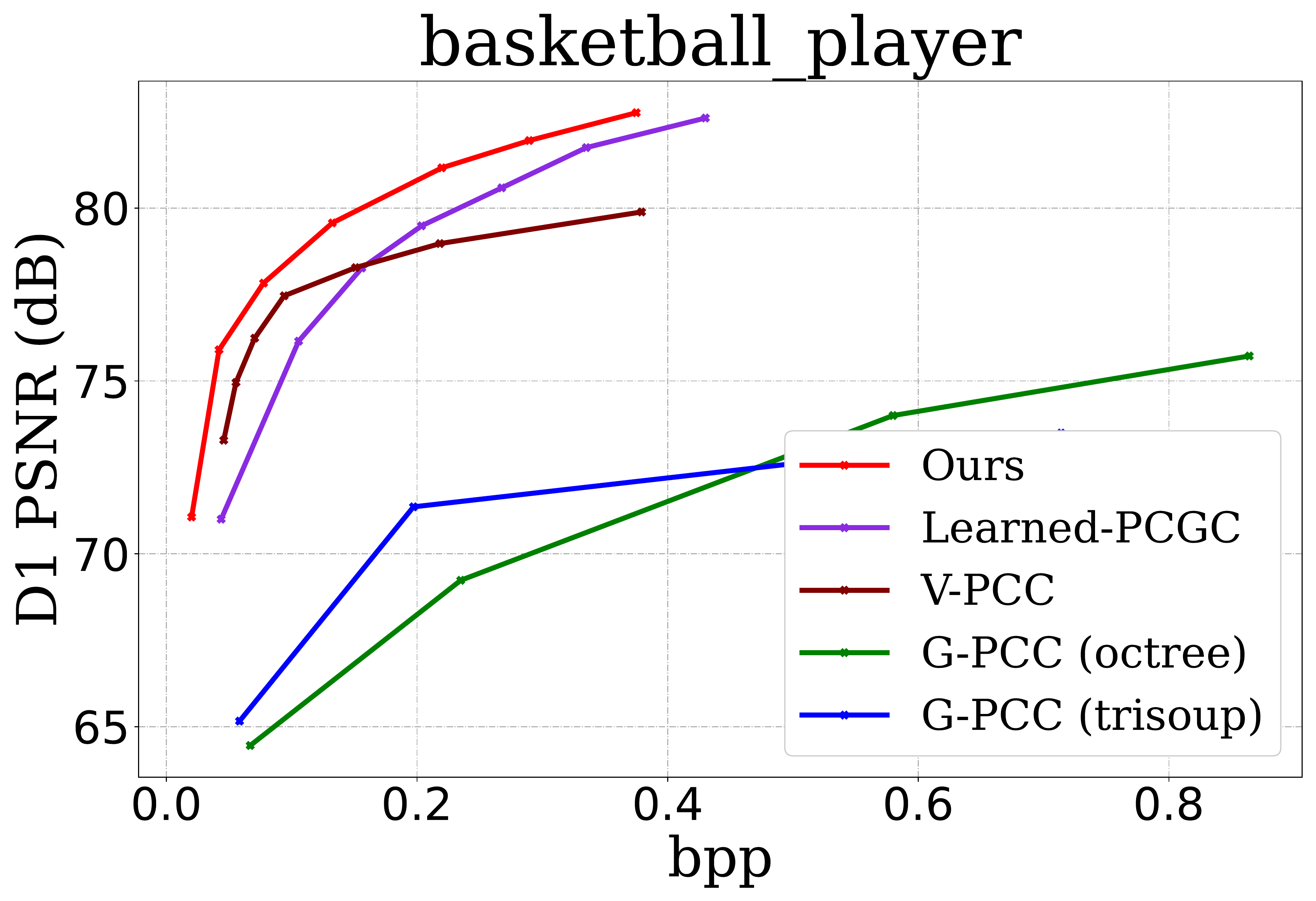}}%
\subfloat{\includegraphics[width=1.42in]{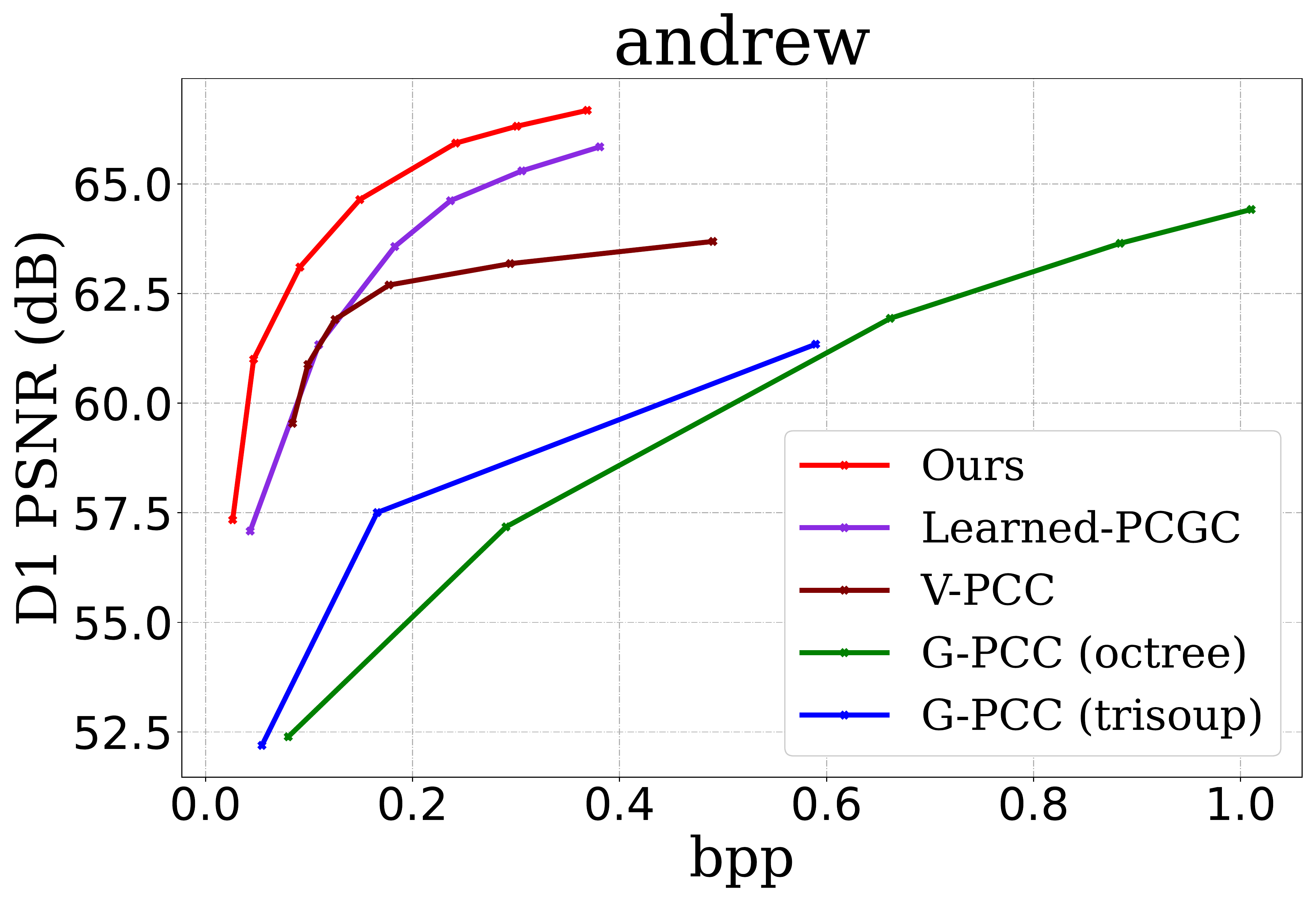}}%
\vspace{-0.1in}
\caption{Rate-distortion efficiency illustration for the state-of-the-arts.
{In each case from left to right, the lossless compressed coordinates $C_Y$ of our method occupy 0.0248, 0.0265, 0.0199, 0.0250 bpp, respectively.}
}
\label{fig:rdcurve}
\end{figure*}

Following the common objective measurement, the bit rate is measured using {\it bits per input point} (bpp), and the distortion is evaluated using point-to-point distance (D1) {and point-to-plane distance (D2)} based mean squared error (MSE) or equivalent Peak Signal-to-Noise Ratio (PSNR). We also plot rate-distortion curves, and calculate the BD-Rate gains as shown in Fig.~\ref{fig:rdcurve} and Table~\ref{table:BDBR}. 
Both MPEG G-PCC and V-PCC are employed for geometry compression. Their reference implementations, i.e., TMC13-v8.1~\cite{tmc13} and TMC2-v7.0~\cite{tmc2} are utilized. We follow the CTC recommendations~\cite{Sebastian2018common} to perform compliant encoding.

\begin{figure*}[htbp]
\vspace{-0.15in}
	\centering
	\includegraphics[width=4.8in]{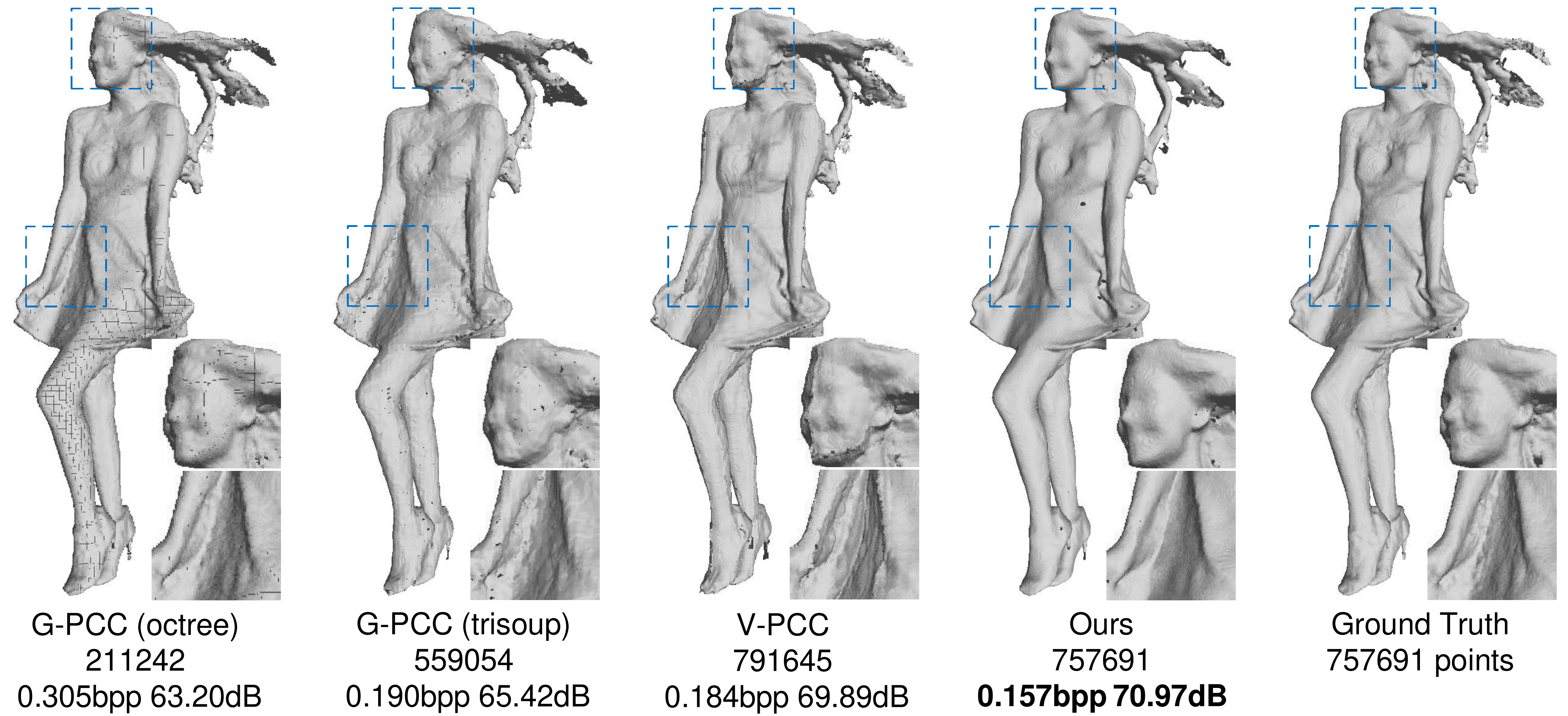}
	\vspace{-0.05in}
	\caption{Visual comparison of ``Redandblack'' for different methods. 
	{In our method, the coordinates and feature attributes occupy 0.026 bpp and 0.130 bpp, respectively}.}
	\label{fig:vis}
	\vspace{-0.1in}
\end{figure*}

As shown in Table~\ref{table:BDBR}, our method achieves average {>80\%} BD-Rate gains against G-PCC (octree) and {>70\%} BD-Rate gains against G-PCC (trisoup).
In the meantime, we obtain {$>$40\%} BD-Rate improvement over V-PCC based geometry compression. {Compared with the Learned-PCGC~\cite{wang2019learned}, our method achieves {$>$ 30\%} BD-Rate gains.}
We also compare different methods subjectively, as visualized in Fig.~\ref{fig:vis}. It is obvious that our method well preserves the geometry details with more smooth and visually appealing reconstruction, while the V-PCC reconstruction has obvious unpleasant seams, and G-PCC reconstruction may lose geometry details.
\begin{table*}[thbp]\small
\vspace{-0.05in}
\caption{BD-Rate gains against the state-of-the-art PCG compression methods using D1 and {D2} distortion measurements.}
\label{table:BDBR}
\centering
\begin{tabular}{c c c c c c c c c}
\hline
& \multicolumn{2}{c}{V-PCC} & \multicolumn{2}{c}{G-PCC (octree)} & \multicolumn{2}{c}{G-PCC (trisoup)} & \multicolumn{2}{c}{Learned-PCGC~\cite{wang2019learned}} \\ 
& D1 & D2 & D1 & D2 & D1 & D2 & D1 & D2 \\ 
\hline
8iVFB & -39.43 & -41.79 & -90.82 & -84.72 & -78.56 & -72.88 & -36.41 & -32.65  \\ 
Owlii & -38.49 & -35.21 & -94.88 & -90.58 & -90.94 & -79.13 & -50.80 & -43.89 \\ 		
MVUB & -60.66 & -53.54 & -90.41 & -83.45  & -87.78 & -79.76 & -46.72 & -39.00  \\ 
{\textbf{Average}} &  \textbf{-47.73} & \textbf{-45.17} &	\textbf{-91.47} & \textbf{-85.38} &  \textbf{-84.72} & \textbf{-76.88} & \textbf{-43.41} & \textbf{-37.44} \\ 
\hline
\end{tabular}
\end{table*}
\vspace{-0.1in}

{\bf Runtime Comparison.}
We compare the runtime of different methods in Table~\ref{table:time}. Tests are benchmarked on a workstation with an Intel Core i7-8700K CPU and an Nvidia GeForce GTX 1070 GPU. Our algorithm prototype is implemented using Python with PyTorch. 
{Because the runtime of G-PCC varies greatly at different bit rates,} for a fair comparison, we collect the encoding and decoding time of the 8iVFB samples coded at the highest bit rate for estimating the worst case.
In Table~\ref{table:time}, our encoding runtime is close to that of the fastest G-PCC octree codec, and much faster than that of V-PCC. 
{For decoding, our solution performs slower than others which mainly because the binary classification based hierarchical reconstruction generates more voxels to be processed and classified. 
{The corresponding floating-point operations (FLOPs) for encoding and decoding are 17.5 and 60 GFLOPs according to our measurements, respectively.} 
We believe runtime can be optimized by simplifying the framework and migrating to the C++ implementation.}
Since all of these prototypes are mostly for performance evaluation purposes, runtime studies only serve as the reference to have a general idea about the complexity of our method.
On the other hand, the GPU running memory of our method is completely affordable, which only requires $\approx$333 MB and 1,273 MB respectively for encoding and decoding, {while Learned-PCGC~\cite{wang2019learned} requires at least 5GB memory.} 
{The number of parameters of our model cost about 778k (about 3.2 MB) which is a fairly small-size network in practices.}
\begin{table*}[htbp]\small
\vspace{-0.05in}
	\caption{Average runtime of different methods using 8iVFB samples.
	}
	\label{table:time}
	\centering
    \begin{tabular}{c c c c c c}
    \hline
    & V-PCC & G-PCC (octree) & G-PCC (trisoup) & Learned-PCGC~\cite{wang2019learned} & Ours \\ 
    \hline
    Enc (s) & 103.45 & \textbf{1.59} & 8.15 & 6.40 & \textbf{1.58} \\
    Dec (s) & \textbf{0.69} & \textbf{0.60}  & 6.57 & 6.75 & 5.40  \\
    \hline
    \end{tabular}
\end{table*}
\vspace{-0.1in}

\section{Discussion}
{\bf Multiscale Reconstruction.}
The experiment result proves that in the runtime and memory cost, our multiscale reconstruction is more efficient than the single scale approach because it prunes out the null voxels from one scale to another.
Geometry reconstruction at the lower scale is upscaled and refined to the next level by classification, in which the complexity is greatly reduced because less intermediate voxels need to be processed. 
{For example, when directly reconstructing the PCG to its largest scale, the number of voxels increases almost $8\times8\times8= 512$-fold because we apply the re-sampling three times. 
While in multiscale reconstruction,  half of the voxels are pruned at each upsampling step, having $8\times\frac{1}{2}\times8\times\frac{1}{2}\times8=128$x increase of voxels from the least scale.
Thus, the single-scale reconstruction requires almost 4$\times$ space to host additional voxels, resulting in around {2.4$\times$ FLOPs increase,} 4$\times$ running memory consumption, and 3$\times$ the overall runtime cost.}


{\bf Adaptive Contexts Conditioned on Priors.}
To better exploit the structural correlations of latent representation,
we model the elements of $\hat{F_{Y}}$ as a conditional Gaussian mixture model (GMM) distribution $\mathcal{N}(\mu_{i}, \sigma_{i}^{2})$, whose parameters (i.e., mean $\mu$ and standard deviation $\sigma$) can be predicted using either autoregressive or hyper priors. 
{Similar to the methods proposed in~\cite{minnen2018joint, balle2018variational} for 2D image, the \textit{autoregressive prior} is realized using 3D masked convolution, where the previous voxels will be used for predicting the context model parameters. The \textit{hyper prior} is realized through downscaling $F_{Y}$.}
Contexts conditioned on either autoregressive prior or hyperprior could offer additional BD-Rate gains.
{In our experiments,} having autoregressive prior and hyper prior achieve averaged -7.28\% and -4.14\% BD-Rate gains on the entire testing dataset, respectively. 
Both of them perform well on Owlii and 8iVFB, but not at same level on MVUB. A possible reason is that the MVUB dataset is more noisy, yielding less correlation between neighbors.  
On the other hand, both of them make the entropy model more complex with the computational penalty. Thus, we defer this topic as our future study for deeper investigation.

\section{Conclusion}
This work has further demonstrated the powerful capacity of deep learning techniques for 3D point cloud geometry compression. We have introduced the multiscale re-sampling to extensively leverage the sparsity nature of the point cloud for compact and efficient feature representation. In the meantime, sparse convolutions are utilized to reduce the space and time complexity when processing the sparse tensor. Such multiscale approach not only improves efficiency, with more than 40\% BD-Rate gains over the state-of-the-art MPEG V-PCC using the common test sequences but also offers reliable and robust model in both training and inference with reduced complexity consumption. As for future studies, an interesting topic is extending this work to include the color attributes. Another exciting avenue is compressing the LiDAR point cloud with the more sparse distribution.


\Section{References}
\bibliographystyle{IEEEbib}
\bibliography{pccbib}

\begin{thebibliography}{10}

\bibitem{Schwarz2019EmergingMS}
Sebastian Schwarz, Marius Preda, Vittorio Baroncini, et~al.,
\newblock ``Emerging mpeg standards for point cloud compression,''
\newblock {\em IEEE Journal on Emerging and Selected Topics in Circuits and
  Systems}, vol. 9, pp. 133--148, 2019.

\bibitem{quach2019learning}
Maurice Quach, Giuseppe Valenzise, and Frederic Dufaux,
\newblock ``Learning convolutional transforms for lossy point cloud geometry
  compression,''
\newblock in {\em 2019 IEEE International Conference on Image Processing
  (ICIP)}. IEEE, 2019, pp. 4320--4324.

\bibitem{guarda2019point}
Andr{\'e}~FR Guarda, Nuno~MM Rodrigues, and Fernando Pereira,
\newblock ``Point cloud coding: Adopting a deep learning-based approach,''
\newblock in {\em 2019 PCS}. IEEE, 2019, pp. 1--5.

\bibitem{wang2019learned}
Jianqiang Wang, Hao Zhu, Zhan Ma, Tong Chen, Haojie Liu, and Qiu Shen,
\newblock ``Learned point cloud geometry compression,''
\newblock {\em arXiv preprint arXiv:1909.12037}, 2019.

\bibitem{Yan2019DeepAL}
Wei Yan, Yiting Shao, Shan Liu, Thomas~H. Li, Zhu Li, and Ge~Li,
\newblock ``Deep autoencoder-based lossy geometry compression for point
  clouds,''
\newblock {\em ArXiv}, vol. abs/1905.03691, 2019.

\bibitem{huang20193d}
Tianxin Huang and Yong Liu,
\newblock ``3d point cloud geometry compression on deep learning,''
\newblock in {\em Proceedings of the 27th ACM International Conference on
  Multimedia}, 2019.

\bibitem{Wen2020ICME}
X.~{Wen}, X.~{Wang}, J.~{Hou}, L.~{Ma}, Y.~{Zhou}, and J.~{Jiang},
\newblock ``Lossy geometry compression of 3d point cloud data via an adaptive
  octree-guided network,''
\newblock in {\em 2020 IEEE International Conference on Multimedia and Expo
  (ICME)}, 2020, pp. 1--6.

\bibitem{Schnabel2006Octree}
Ruwen Schnabel and Reinhard Klein,
\newblock ``Octree-based point-cloud compression,''
\newblock in {\em Eurographics}, 2006.

\bibitem{choy20194d}
Christopher Choy, JunYoung Gwak, and Silvio Savarese,
\newblock ``4d spatio-temporal convnets: Minkowski convolutional neural
  networks,''
\newblock in {\em Proceedings of the IEEE Conference on Computer Vision and
  Pattern Recognition}, 2019, pp. 3075--3084.

\bibitem{Sullivan2013Overview}
Gary~J. Sullivan, J.~R. Ohm, and Thomas Wiegand,
\newblock ``Overview of the high efficiency video coding (hevc) standard,''
\newblock {\em IEEE Transactions on Circuits \& Systems for Video Technology},
  vol. 22, no. 12, pp. 1649--1668, 2013.

\bibitem{minnen2018joint}
David Minnen, Johannes Ball{\'e}, and George~D Toderici,
\newblock ``Joint autoregressive and hierarchical priors for learned image
  compression,''
\newblock in {\em Advances in Neural Information Processing Systems}, 2018, pp.
  10771--10780.

\bibitem{wang2017cnn}
Peng-Shuai Wang, Yang Liu, Yu-Xiao Guo, Chun-Yu Sun, and Xin Tong,
\newblock ``O-cnn: Octree-based convolutional neural networks for 3d shape
  analysis,''
\newblock {\em ACM Transactions on Graphics (TOG)}, vol. 36, no. 4, pp. 1--11,
  2017.

\bibitem{graham20183d}
Benjamin Graham, Martin Engelcke, and Laurens van~der Maaten,
\newblock ``3d semantic segmentation with submanifold sparse convolutional
  networks,''
\newblock in {\em Proceedings of the IEEE conference on computer vision and
  pattern recognition}, 2018, pp. 9224--9232.

\bibitem{Graham2018UnsupervisedLW}
Benjamin Graham,
\newblock ``Unsupervised learning with sparse space-and-time autoencoders,''
\newblock {\em ArXiv}, vol. abs/1811.10355, 2018.

\bibitem{Szegedy2017Inceptionv4IA}
Christian Szegedy, S.~Ioffe, V.~Vanhoucke, and Alexander~Amir Alemi,
\newblock ``Inception-v4, inception-resnet and the impact of residual
  connections on learning,''
\newblock in {\em AAAI}, 2017.

\bibitem{Tang2020DeepIV}
Danhang Tang, Saurabh Singh, P.~Chou, Christian H{\"a}ne, M.~Dou, S.~Fanello,
  et~al.,
\newblock ``Deep implicit volume compression,''
\newblock {\em 2020 IEEE/CVF Conference on Computer Vision and Pattern
  Recognition (CVPR)}, pp. 1290--1300, 2020.

\bibitem{tmc13}
MPEG,
\newblock ``Mpeg-pcc-tmc13,''
  \url{https://github.com/MPEGGroup/mpeg-pcc-tmc13},
\newblock Accessed: 2020.

\bibitem{balle2018variational}
Johannes Ballé, David Minnen, Saurabh Singh, Sung~Jin Hwang, and Nick
  Johnston,
\newblock ``Variational image compression with a scale hyperprior,''
\newblock in {\em International Conference on Learning Representations}, 2018.

\bibitem{chang2015shapenet}
Angel~X Chang, Thomas Funkhouser, Leonidas Guibas, et~al.,
\newblock ``Shapenet: An information-rich 3d model repository,''
\newblock {\em arXiv preprint arXiv:1512.03012}, 2015.

\bibitem{8i20178i}
Eugene d'Eon, Bob Harrison, Taos Myers, and Philip A.~Chou,
\newblock ``8i voxelized full bodies - a voxelized point cloud dataset,''
\newblock {\em ISO/IEC JTC1/SC29 Joint WG11/WG1 (MPEG/JPEG) input document
  m38673/M72012}, May 2016.

\bibitem{xu2017owlii}
Xu~Yi, Lu~Yao, and Wen Ziyu,
\newblock ``Owlii dynamic human mesh sequence dataset,''
\newblock {\em ISO/IEC JTC1/SC29/WG11 (MPEG/JPEG) m41658}, October 2017.

\bibitem{microsoft2019microsoft}
Loop Charles, Cai Qin, O.Escolano Sergio, and A.~Chou Philip,
\newblock ``Microsoft voxelized upper bodies - a voxelized point cloud
  dataset,''
\newblock {\em ISO/IEC JTC1/SC29 Joint WG11/WG1 (MPEG/JPEG) input document
  m38673/M72012}, May 2016.

\bibitem{Sebastian2018common}
Sebastian Schwarz, Philip~A. Chou, and Indranil Sinharoy,
\newblock ``Common test conditions for point cloud compression,''
\newblock {\em ISO/IEC JTC1/SC29/WG11 N18474}, 2018.

\bibitem{JPEG2019CTC}
JPEG~Pleno PCC,
\newblock ``Jpeg pleno point cloud coding common test conditions,''
\newblock {\em JPEG (ISO/IEC JTC 1/SC 29/WG1)}, 2019.

\bibitem{tmc2}
MPEG,
\newblock ``Mpeg-pcc-tmc2,'' \url{https://github.com/MPEGGroup/mpeg-pcc-tmc2},
\newblock Accessed: 2020.

\end{thebibliography}

\end{document}